\documentclass[sigconf]{acmart}

\usepackage{multirow}
\usepackage{booktabs}    %
\usepackage[table]{xcolor}%

\usepackage{pdfpages} %

\AtBeginDocument{%
  }
\newcommand{\TopRule}{\toprule}

\copyrightyear{2026}
\acmYear{2026}
\setcopyright{acmlicensed}\acmConference[MM '26]{Proceedings of the 34nd ACM International Conference on Multimedia}{November 10-14, 2026}{Rio de Janeiro, Brazil}
\acmBooktitle{Proceedings of the 34nd ACM International Conference on Multimedia (MM '26), November 10-14, 2026, Rio de Janeiro, Brazil}
\acmDOI{xxxx}
\acmISBN{xxxx}

\begin{document}

\title{Dual-Stream Decoupled Learning for Temporal Consistency and Speaker Interaction in AVSD}



\author{Junhao Xiao}
\authornote{Equal contribution.}
\email{xiaojunhao431@mails.ccnu.edu.cn}
\affiliation{%
  \institution{CCNU \& Kuaishou}
  \city{Wuhan}
  \country{China}
}

\author{Shun Feng}
\authornotemark[1]
\email{fengshun@mails.ccnu.edu.cn}
\affiliation{%
  \institution{CCNU}
  \city{Wuhan}
  \country{China}
}

\author{Zhiyu Wu}
\email{wuzy24@m.fudan.edu.cn}
\affiliation{%
  \institution{FDU}
  \city{Shanghai}
  \country{China}
}

\author{Jinghan Yu}
\email{jinghanyu0917@gmail.com}
\affiliation{%
  \institution{HUST}
  \city{Wuhan}
  \country{China}
}

\author{Haibiao Yao}
\email{yaohb@mail.ustc.edu.cn}
\affiliation{%
  \institution{USTC}
  \city{Hefei}
  \country{China}
}

\author{Zhiyuan Ma}
\email{mzyth@hust.edu.cn}
\affiliation{%
  \institution{HUST}
  \city{Wuhan}
  \country{China}
}

\author{Jianjun Li}
\email{jianjunli@hust.edu.cn}
\affiliation{%
  \institution{HUST}
  \city{Wuhan}
  \country{China}
}

\author{Youjun Bao}
\email{baoyoujun@kuaishou.com}
\affiliation{%
  \institution{Kuaishou}
  \city{Beijing}
  \country{China}
}

\author{Yi Chen}
\authornote{Corresponding author.}
\email{chenyi30@mail.ccnu.edu.cn}
\affiliation{%
  \institution{CCNU}
  \city{Wuhan}
  \country{China}
}

\renewcommand{\shortauthors}{Xiao, et al.}


\begin{abstract}
Audio-Visual Speaker Detection (AVSD) hinges on modeling both individual temporal continuity and inter-personal social context. Existing coupled architectures struggle to reconcile these tasks in shared representation spaces due to conflicting inductive biases: temporal modeling favors low-frequency smoothness, while inter-personal interaction requires high-frequency discriminability. We propose D$^2$Stream, a decoupled dual-stream framework that explicitly isolates these functionalities into parallel, task-specific branches. Specifically, the Intra-speaker Temporal Continuity (ITC) stream captures longitudinal stability, whereas the Inter-personal Social Relation (ISR) stream models transversal social cues. Quantitative gradient analysis reveals an evolutionary divergence in update directions, stabilizing at $86.1^\circ$, which confirms the inherent task conflict and the effectiveness of our structural decoupling. D$^2$Stream breaks the long-standing performance plateau, achieving a state-of-the-art 95.6\% mAP on AVA-ActiveSpeaker and superior generalization on Columbia ASD, all within a lightweight and efficient design. \textcolor{blue}{Code is publicly available at \href{https://anony-xyz.github.io/}{\texttt{project page}}}.

\end{abstract}



\maketitle

\begin{figure}[t]
    \centering
  \includegraphics[width=0.45\textwidth]{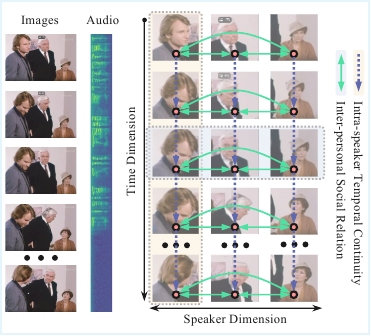}
    \caption{AVSD is decomposed into two orthogonal axes: (1) Intra-speaker Temporal Continuity (ITC) (vertical axis, blue), which captures longitudinal stability and temporal smoothness for individual speakers; and (2) Inter-personal Social Relation (ISR) (horizontal axis, green), which models transversal social cues and contrastive context among multiple individuals.}
    \label{fig:syt}
\end{figure}

\section{Introduction}
Audio-Visual Speaker Detection (AVSD) aims to accurately localize active speakers from complex audio-visual streams, serving as a fundamental building block for intelligent conferencing, scene understanding, and human-computer interaction~\cite{ASC, EASEE, Real-Time, OpticalFlow}. Although this task is often treated as an alignment problem between audio and visual signals~\cite{audio-visual}, in real-world social scenarios, speaking behavior is not only reflected by lip-audio synchronization but is also implicitly embedded in complex multi-person interactions. Particularly under conditions such as lip occlusion, low visual quality, or non-verbal interference, identifying the active speaker often relies on recognizing the ``interaction focus'', i.e., inferring the speaker indirectly by observing gaze, reactions, and contextual dependencies among individuals~\cite{S-VVAD}.

As illustrated in Fig.~\ref{fig:syt}, AVSD inherently involves two distinct modeling dimensions. The first is \textbf{Intra-speaker Temporal Continuity (ITC)} modeling, which operates along the temporal axis to capture the behavior of the same speaker over time, ensuring temporal consistency between audio and visual signals. The second is \textbf{Inter-personal Social Relation (ISR)} modeling, which operates along the speaker (spatial) axis to capture interactions among different individuals within the same frame, leveraging group context to facilitate speaker identification. For example, when a person's mouth is occluded, the model can infer the speaker by observing others' gaze or reactions (e.g., head turns, attentive listening). These two types of information correspond to ``temporal continuity evidence'' and ``spatial contrastive evidence'', respectively.

However, existing methods typically adopt a single coupled architecture that jointly models both types of interactions within a shared representation space~\cite{ASC,EASEE,ASDNet}. Such designs implicitly assume that these two modeling capabilities can be effectively captured by a unified set of feature parameters. From the perspective of representation learning and inductive bias, this assumption is fundamentally flawed. ITC modeling favors low-frequency smoothness along the temporal axis to maintain stable audio-visual correspondence, whereas ISR modeling relies on capturing subtle contrastive cues within groups, emphasizing high-frequency discriminability in the spatial domain. When both are forced to share parameters, their optimization objectives inevitably interfere with each other.

\begin{figure}[t]
    \centering
  \includegraphics[width=0.5\textwidth]{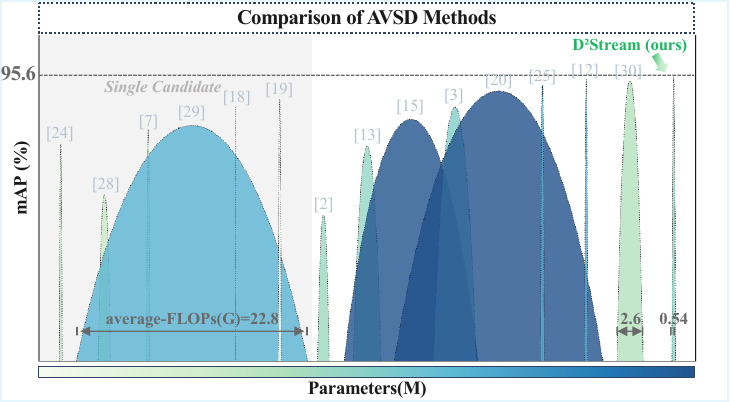}
    \caption{Visualization of AVSD methods on AVA-ActiveSpeaker. Each shape encodes three factors simultaneously: height indicates mAP (performance), width reflects FLOPs (computational cost), and color depth represents parameter scale (darker denotes larger models). Our D²Stream achieves superior performance with lower complexity compared to prior methods.}
    \Description{A comparison chart of methods on AVA-ActiveSpeaker showing mAP, FLOPs, and parameter count.}
    \label{fig:pkt}
\end{figure}

Motivated by this observation, we propose \textbf{D$^{2}$Stream}, a simple yet elegant decoupled dual-stream framework. Instead of improving performance by stacking contextual dimensions, our key idea is to explicitly decouple these two inherently orthogonal modeling capabilities through architectural parallelism, allowing each to evolve within its own dedicated space. Specifically, built upon a multimodal feature backbone, we design two functionally specialized interaction streams: the \textbf{ITC-Stream}, which focuses on modeling longitudinal consistency at the individual level using long-range temporal modeling to capture low-frequency smooth dynamics, ensuring robust audio-visual alignment; and the \textbf{ISR-Stream}, which focuses on modeling horizontal relationships across multiple individuals by extracting high-frequency discriminative features from social context, enabling precise identification of the interaction focus from global relations. The two streams maintain structural symmetry and simplicity, differing only in the dimension along which attention is applied, thereby injecting distinct inductive biases.

To validate our theoretical hypothesis, we conduct a quantitative gradient analysis of the two streams. Experimental results show that the gradient directions of the two branches on shared parameters progressively diverge during training, with the average angle increasing from an initial $63.9^\circ$ to a stable $86.1^\circ$. This near-orthogonality at the gradient level provides direct evidence of the heterogeneous feature requirements between single-speaker temporal modeling and multi-person spatial modeling. Furthermore, architectural comparisons demonstrate that, under the same parameter budget, the parallel decoupled design achieves a notable improvement of 0.5 mAP over its serially coupled counterpart, further confirming the effectiveness of decoupling.


On the authoritative benchmark dataset AVA-ActiveSpeaker~\cite{AVA}, as shown in Fig.~\ref{fig:syt}, D$^{2}$Stream establishes a new state-of-the-art performance of 95.6 mAP with a lightweight and simple architecture, breaking the stagnation observed since 2024~\cite{TalkNCE}. Moreover, its strong performance on the Columbia ASD dataset~\cite{Columbia} further demonstrates its generalization capability.

Our contributions can be summarized as follows:
\begin{itemize}
    \item \textbf{Structural Insights:} We decompose AVSD into two orthogonal dimensions, longitudinal ITC and transverse ISR, exposing the bottleneck in coupled architectures due to conflicting inductive biases.
    
    \item \textbf{Methodological Innovation:} We propose a lightweight D$^2$Stream framework with explicit parallel decoupling, using dimension-oriented attention to replace redundant stacking and unify accuracy and efficiency.
    
    \item \textbf{Empirical \& Theoretical Validation:} We validate the necessity of decoupling via gradient angle analysis, and achieve new SOTA on AVA-ActiveSpeaker after nearly two years of stagnation, with strong cross-scenario generalization.
\end{itemize}

\section{Related Work}

The core challenge of AVSD lies in resolving semantic ambiguity in multi-speaker scenarios from complex audio-visual streams. Existing research has evolved from single-speaker temporal modeling to multi-person social relation modeling, and further to unified context integration.

\subsection{Single-person Temporal Modeling and Early Exploration}

Early ASD methods primarily adopt a single-person perspective, emphasizing the importance of temporal modeling for speaker identification. Representative works such as TalkNet~\cite{TalkNet} and Light-ASD~\cite{Light-ASD} capture long-range audio-visual correspondence via self-attention mechanisms and lightweight sequence models, respectively. While these methods validate the effectiveness of temporal consistency modeling, they typically process each candidate face track independently, ignoring interactions among individuals within the same frame. In densely populated conversations or visually noisy scenarios, such paradigms often fail to suppress false activations of non-speakers.

Recent works further improve robustness from the perspective of modality modeling. For example, SCAN~\cite{SCAN} leverages reference audio for frame-level contrastive modeling to strengthen speaker identity constraints; GateFusion~\cite{GateFusion} introduces a hierarchical gated fusion architecture to model fine-grained audio-visual dependencies via cross-level bidirectional modulation; LASER~\cite{LASER} explicitly incorporates lip keypoints as structural priors to guide the model toward regions highly correlated with speech production. Additionally, works such as LR-ASD~\cite{LR-ASD} focus on efficiency, reducing computational cost through lightweight designs while maintaining performance. Although these methods enhance discriminability through cross-modal interaction or local structural modeling, they still primarily rely on single-person temporal cues.

\subsection{Multi-person Relation Modeling and Graph-based Methods}
To address the limitations of single-person modeling, subsequent works introduce explicit multi-person relation modeling mechanisms. Graph-based approaches construct connections among candidates to capture social context. For instance, MAAS~\cite{MAAS}, EASEE~\cite{EASEE}, and SPELL~\cite{SPELL+} model multi-speaker interactions via local graph structures, alternating message passing, and long spatio-temporal graphs, respectively; AFs-Net~\cite{AFs-Net} further enhances intra-frame alignment through graph attention mechanisms. Despite their effectiveness in modeling inter-person relations, graph-based methods incur substantial computational and memory overhead due to explicit graph construction and multi-round message passing.

\subsection{Unified Context Modeling and Sequential Architectures}
More recently, a line of work attempts to jointly model multiple contextual cues within a unified framework, capturing all candidates in a scene from a global perspective. ASC~\cite{ASC} introduces speaker context by encoding short-term audio-visual features with a dual-stream structure, followed by self-attention and sequence models to capture pairwise relations and long-term temporal interactions among speakers. UniCon~\cite{Unicon} further constructs a unified context framework that integrates spatial positional information, inter-speaker contrastive relations, and temporal continuity, jointly optimizing all candidates in an end-to-end manner. LoCoNet~\cite{LocoNet} proposes a long-short context network, employing self-attention to model long-term temporal dependencies for individual speakers while using convolutional blocks to capture short-term local interactions among multiple individuals. Building upon LoCoNet, TalkNCE~\cite{TalkNet} introduces a speaker-aware contrastive loss to further enhance representation learning, improving the detection performance of LoCoNet to 95.5 mAP on the AVA-ActiveSpeaker dataset~\cite{AVA} and maintaining a leading position in subsequent works. This prolonged stagnation in performance growth suggests that existing modeling paradigms have reached a bottleneck.

\begin{figure}[t]
  \centering
  \includegraphics[width=0.48\textwidth]{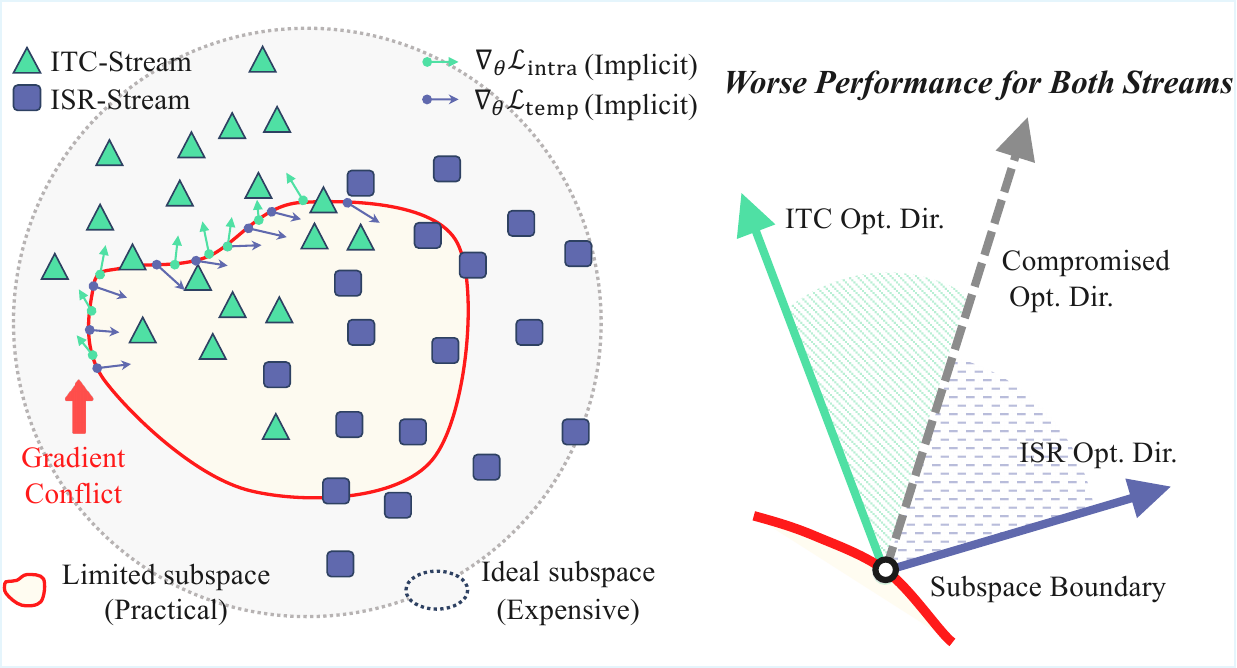}
  \caption{Implicit gradient conflict and representation compromise in a coupled architecture.}
  \label{fig:yqt}
\end{figure}

\section{Motivation and Key Assumptions}
\label{sec:assum}
Existing methods usually model ITC (Intra-speaker Temporal Continuity) and ISR (Inter-personal Social Relation) in a single coupled representation space. This design assumes that both modeling objectives can be captured by the same set of parameters. However, from the perspective of inductive bias in representation learning, their optimization goals on the feature manifold are fundamentally conflicting.

The first objective is temporal continuity for ITC. It aims to maintain long-range coherence between audio and the corresponding visual signal. This requires reducing feature variance of the same person across adjacent frames, leading to low-frequency smoothness. Its implicit objective can be formulated as minimizing the temporal continuity loss:
\begin{equation}
\mathcal{L}_{\text{intra}} \propto \sum_{t} \|f_{i, t+1} - f_{i, t}\|^2 \rightarrow \min
\end{equation}

The second objective is spatial discriminability for ISR. It aims to distinguish different individuals at the same time and model their relations. This requires enlarging the feature distance between different persons, leading to high-frequency contrast. Its implicit objective can be formulated as maximizing inter-person differences:
\begin{equation}
\mathcal{L}_{\text{inter}} \propto \sum_{i \neq j} \|f_{i, t} - f_{j, t}\|^2 \rightarrow \max
\end{equation}

As shown in Fig.~\ref{fig:yqt}, these two objectives occupy different optimal regions in the latent representation space (blue squares for ITC and green triangles for ISR). In a coupled architecture, the model capacity is limited. The shared parameter space leads to gradient conflict. When minimizing $\mathcal{L}_{\text{intra}}$ and maximizing $\mathcal{L}_{\text{inter}}$, the gradients $\nabla_{\theta}\mathcal{L}_{\text{intra}}$ and $\nabla_{\theta}\mathcal{L}_{\text{inter}}$ often point to different directions. Under shared parameters, the model is forced to follow a compromised optimal direction.

This leads to a see-saw effect. Optimization for temporal consistency introduces noise for spatial discrimination, and vice versa. Under limited model capacity, the shared representation space cannot fit both objectives well~\cite{AgenticRL}. This limits performance in complex multi-speaker scenarios.

Based on this analysis, we propose the following assumption: decoupling ITC and ISR into separate subspaces and optimizing them independently can reduce gradient conflict and improve overall performance.

\section{Proposed Method}
\begin{figure*}[t]
  \centering
  \includegraphics[width=0.95\textwidth]{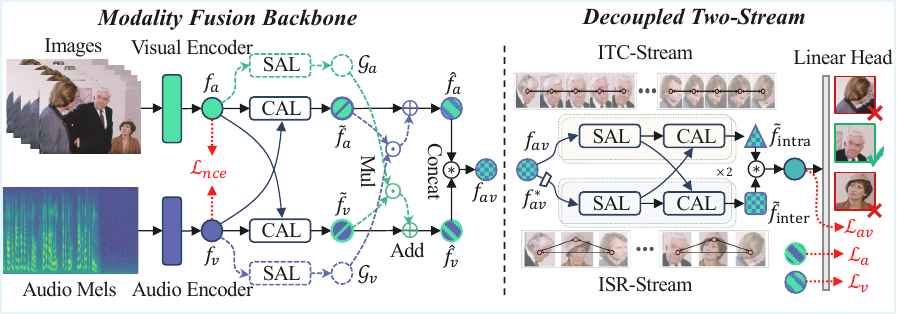}
  \caption{Overview of the decoupled dual-stream framework.}
  \Description{Placeholder for Figure 3 showing the framework of D2Stream.}
  \label{fig:framework-overview}
\end{figure*}

Guided by the the analysis in Section~\ref{sec:assum}, we design a simple decoupled dual-stream framework D$^2$Stream. The overall architecture is shown in Fig.~\ref{fig:framework-overview}. It consists of a multimodal fusion backbone and two decoupled streams: ITC-Stream and ISR-Stream.

\subsection{Basic Attention Layer}

Our model is simple and built on two types of interaction layers.

\paragraph{Self-Attention Interaction Layer (SAL)}
SAL consists of multi-head self-attention (MHSA), a feed-forward network (MLP), and layer normalization (LN) with residual connections. For the $\ell$-th layer, the input $X^{(\ell)}$ is updated as:
\begin{equation}
Z^{(\ell)} = \operatorname{LN}\big(X^{(\ell)} + \operatorname{MHSA}(X^{(\ell)})\big)
\end{equation}
\begin{equation}
X^{(\ell+1)} = \operatorname{LN}\big(Z^{(\ell)} + \operatorname{MLP}(Z^{(\ell)})\big)
\end{equation}

\paragraph{Cross-Attention Interaction Layer (CAL)}
CAL is used for cross-modal interaction. It replaces MHSA with multi-head cross-attention (MHCA). It models the relation between a query modality $X$ and a key and value modality $Y$:
\begin{equation}
Z^{(\ell)} = \operatorname{LN}\big(X^{(\ell)} + \operatorname{MHCA}(X^{(\ell)}, Y^{(\ell)}, Y^{(\ell)})\big)
\end{equation}
\begin{equation}
X^{(\ell+1)} = \operatorname{LN}\big(Z^{(\ell)} + \operatorname{MLP}(Z^{(\ell)})\big)
\end{equation}

\subsection{Multimodal Fusion Backbone}

We design a simple backbone to fuse audio and visual information.
Given a video clip with $T$ frames and $S$ persons per frame, the visual input is face crops: $V \in \mathbb{R}^{S \times T \times H \times W \times 1}$. The audio input is the mel-spectrogram: $A \in \mathbb{R}^{4T \times M}$.

We extract embeddings using a visual encoder $G_v(\cdot)$ and an audio encoder $G_a(\cdot)$:
\begin{equation}
f_v = G_v(V), \quad f_a = \operatorname{Repeat}_S(G_a(A)) \in \mathbb{R}^{S \times T \times C}
\end{equation}
We align the two modalities using bidirectional cross-attention:
\begin{equation}
\tilde{f}_a = \operatorname{CAL}(f_a, f_v, f_v), \quad
\tilde{f}_v = \operatorname{CAL}(f_v, f_a, f_a)
\end{equation}
Cross-attention captures similarity between features. However, it may be sensitive to noise or irrelevant persons. To address this, we introduce a context-aware gating mechanism. It uses intra-modal context to control cross-modal feature injection.
We first extract context features using SAL and generate gating vectors:
\begin{equation}
\mathcal{G}_a = \sigma(\operatorname{SAL}(f_a)), \quad
\mathcal{G}_v = \sigma(\operatorname{SAL}(f_v))
\end{equation}
where $\sigma(\cdot)$ is the sigmoid function.
We then modulate the aligned features:
\begin{equation}
\hat{f}_a = \tilde{f}_a + \tilde{f}_a \odot \mathcal{G}_v, \quad
\hat{f}_v = \tilde{f}_v + \tilde{f}_v \odot \mathcal{G}_a
\end{equation}
This design uses confidence from the other modality to control feature flow. For example, when audio is reliable, $\mathcal{G}_a$ enhances relevant visual features. When noise exists, the gate suppresses incorrect alignment.
Finally, we concatenate the features:
\begin{equation}
f_{av} = \circledast(\hat{f}_a, \hat{f}_v) \in \mathbb{R}^{S \times T \times 2C}
\end{equation}
where $\circledast(\cdot,\cdot)$ denotes concatenation.

\subsection{ITC-Stream}

The multimodal fused representation $f_{av}$ encodes initial audio-visual correspondence. However, it is still local and frame-wise. It lacks global temporal consistency for speaking behavior. As discussed in Section~\ref{sec:assum}, ITC (Intra-speaker Temporal Continuity) aims to learn a low-frequency and smooth dynamic function. This function models the continuous evolution of audio-visual signals over time.

We explicitly construct the ITC-Stream. It models temporal dependencies in an independent subspace. This design avoids gradient interference from ISR (Inter-personal Social Relation) modeling.
Given the fused feature $f_{av} \in \mathbb{R}^{S \times T \times 2C}$, 
we reshape it by merging the speaker dimension into the batch dimension:
\begin{equation}
X = \operatorname{Reshape}(f_{av}) \in \mathbb{R}^{(S) \times T \times 2C}.
\end{equation}
In this representation, each sequence corresponds to the full temporal trajectory of a single speaker. We then apply the SAL:
\begin{equation}
f_{\text{intra}} = \operatorname{SAL}(X).
\end{equation}
In this temporal dimension, self-attention aggregates global context across frames. It compensates for missing information caused by occlusion or transient noise. The output feature captures smooth and consistent temporal dynamics of speaking behavior.

\subsection{ISR-Stream}

Speaker detection in multi-person scenarios often depends on relative relations among individuals within the same frame, such as gaze, response, or semantic focus. This process is essentially a structured relational reasoning problem. Its goal is to enhance discriminability and contrast among individuals.

We design the ISR-Stream to model intra-frame relations in an independent subspace. This avoids interference from ITC (Intra-speaker Temporal Continuity) modeling.
Given the fused representation $f_{av} \in \mathbb{R}^{S \times T \times 2C}$,
we first swap the temporal and speaker dimensions: $f_{av}^{*} \in \mathbb{R}^{T \times S \times 2C}$.
We then reshape the tensor by merging the temporal dimension into the batch dimension. In this form, the set of speakers within each frame is treated as a sequence. To enhance identity discrimination, we introduce learnable speaker embeddings: $\mathbf{E}_{\text{speaker}} \in \mathbb{R}^{S \times 2C}$.
The input to the ISR-Stream becomes:
\begin{equation}
Y = \operatorname{Reshape}(f_{av}^{*}) + \mathbf{E}_{\text{speaker}} \in \mathbb{R}^{(T) \times S \times 2C}.
\end{equation}
We then apply the SAL:
\begin{equation}
f_{\text{inter}} = \operatorname{SAL}(Y).
\end{equation}
In this spatial dimension, self-attention builds full connections among speakers. It explicitly models interaction strength, such as gaze or response relations. The resulting features use contextual cues to support accurate identification of the active speaker.

\subsection{Decoupled Dual-Stream Interaction}

Through the above design, we obtain two streams with different inductive biases: ITC-Stream focuses on low-frequency and smooth temporal modeling, while ISR-Stream focuses on high-frequency and structured relational modeling. However, fully independent modeling leads to information separation. It cannot support joint reasoning between audio-visual alignment and social relations.

To address this issue, we introduce a symmetric cross-stream attention mechanism.
First, we align the feature dimensions of the two streams. The ITC-Stream output $f_{\text{intra}} \in \mathbb{R}^{S \times T \times 2C}$ 
and the ISR-Stream output
$f_{\text{inter}} \in \mathbb{R}^{S \times T \times 2C}$
are transformed into the same shape for interaction.
We then apply CAL for information exchange. In each step, one stream serves as the query, and the other serves as key and value:
\begin{equation}
\tilde{f}_{\text{intra}} = \operatorname{CAL}(f_{\text{intra}}, f_{\text{inter}}, f_{\text{inter}})
\end{equation}
\begin{equation}
\tilde{f}_{\text{inter}} = \operatorname{CAL}(f_{\text{inter}}, f_{\text{intra}}, f_{\text{intra}})
\end{equation}
After cross-attention, each stream is further refined by the SAL. The same cross-stream interaction is applied again to enhance feature exchange.
Finally, the two streams are combined and passed into a prediction head. The head is implemented as a single linear layer. It outputs the final speaker prediction results.

\subsection{Loss Function}

During training, we adopt a multi-loss optimization strategy. It ensures strong unimodal discrimination and improves audio-visual semantic alignment.

For the $s$-th speaker at the $t$-th frame, we denote the ground-truth label as $\mathbf{y}_{s,t} \in \{0,1\}$ and the valid sample mask as $\mathbf{m}_{s,t}$. We compute masked cross-entropy losses for the fusion branch prediction $\mathbf{P}_{av}$ and the auxiliary unimodal predictions $\mathbf{P}_{a}$ and $\mathbf{P}_{v}$.
Taking the audio-visual fusion loss $\mathcal{L}_{av}$ as an example:
\begin{equation}
\mathcal{L}_{av} = -\frac{1}{\sum_{s,t}\mathbf{m}_{s,t}}
\sum_{s,t} \mathbf{m}_{s,t}
\Big[
\mathbf{y}_{s,t}\log\mathbf{P}_{av}^{(s,t)}[1]
+ (1-\mathbf{y}_{s,t})\log\mathbf{P}_{av}^{(s,t)}[0]
\Big],
\end{equation}
The auxiliary losses $\mathcal{L}_{a}$ and $\mathcal{L}_{v}$ follow the same formulation. They are computed using $\mathbf{P}_{a}$ and $\mathbf{P}_{v}$ with the same ground-truth labels $\mathbf{y}_{s,t}$.

To explicitly enforce audio-visual consistency at the feature level, we introduce a contrastive loss~\cite{TalkNCE} on active speakers. We define the active set as $\mathcal{K} = \{(s,t) \mid \mathbf{y}_{s,t} = 1\}$
We maximize the agreement between audio and visual features of the same speaker, and minimize the association across different active speakers:
\begin{equation}
\mathcal{L}_{\text{nce}} = -\frac{1}{|\mathcal{K}|} 
\sum_{i \in \mathcal{K}} 
\log 
\frac{
\exp\left(\langle z_v^{(i)}, z_a^{(i)} \rangle / \tau \right)
}{
\sum_{j \in \mathcal{K}, j \neq i} 
\exp\left(\langle z_v^{(i)}, z_a^{(j)} \rangle / \tau \right)
}
\end{equation}
Here, $\tau = 0.07$ is the temperature parameter, and $\langle \cdot, \cdot \rangle$ denotes the inner product between feature vectors.

The final loss is a weighted sum of all components:
\begin{equation}
\mathcal{L}_{\text{total}} = \mathcal{L}_{av} + \lambda_1 (\mathcal{L}_{a} + \mathcal{L}_{v}) + \lambda_2 \mathcal{L}_{\text{nce}}
\end{equation}
In our experiments, we set $\lambda_1 = 0.4$ and $\lambda_2 = 0.3$. The model is trained end-to-end using backpropagation.

\section{Gradient Properties and Serial-Parallel Architecture Analysis}

\subsection{Gradient Angle Analysis of Dual Streams}
\label{sec:gaa}

\begin{figure*}[t]
  \centering
  \includegraphics[width=0.80\textwidth]{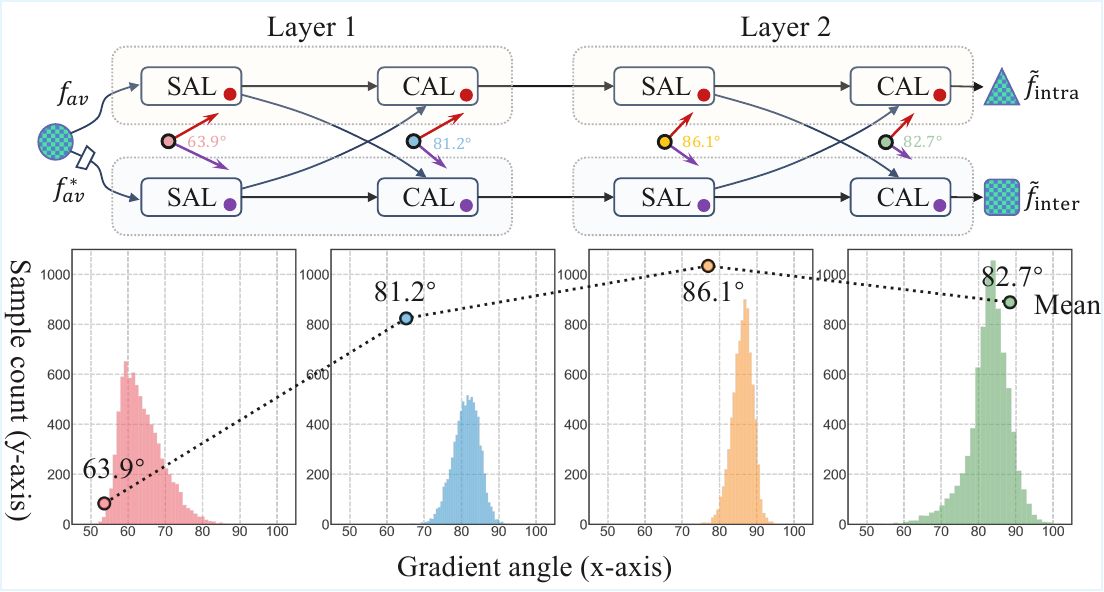}
  \caption{Visualization of gradient-angle evolution in the dual-stream architecture.}
  \Description{Placeholder for Figure 4 showing gradient direction evolution and angle histograms.}
  \label{fig:gradient-angle}
\end{figure*}

To validate the hypothesis in Section~\ref{sec:assum} that gradient conflict exists in a shared representation space, and to analyze the optimization behavior of the proposed decoupled structure, we conduct a quantitative analysis of gradient directions during training.

Let the gradients of ITC-Stream (Intra-speaker Temporal Continuity) and ISR-Stream (Inter-personal Social Relation) be denoted as $g_{\text{intra}}$ and $g_{\text{inter}}$, respectively. Their directional relationship is measured by cosine similarity:
\begin{equation}
\cos(\theta) = \frac{\langle g_{\text{intra}}, g_{\text{inter}} \rangle}{\|g_{\text{intra}}\| \cdot \|g_{\text{inter}}\|}
\end{equation}
Here, $\theta$ denotes the angle between the two gradient directions. We compute this metric over all 8015 samples in the AVA-ActiveSpeaker validation set.

\begin{table}[t]
  \centering
    \caption{Comparison between parallel dual-stream and sequential dual-stream designs.}
  \begin{tabular}{@{}lccc@{}}
    \toprule
    Ablation Setting & Params(M)$\downarrow$ & FLOPs(G)$\downarrow$ & mAP(\%) \\
    \midrule
    Sequential & $22.23$ & $0.54$ & $95.1$ \\
    Parallel   & $24.34$ & $0.54$ & $95.6$ \\
    \bottomrule
  \end{tabular}
  \label{tab:ablation_parallel}
\end{table}

The measurement is conducted at four key stages, as illustrated in Fig.~\ref{fig:gradient-angle}. These stages include after the first SAL, after the first CAL, after the second SAL, and after the second CAL. The corresponding average angles are $63.9^\circ$, $81.2^\circ$, $86.1^\circ$, and $82.7^\circ$, respectively. This trend reveals a progressive functional separation inside the model. At the initial stage ($63.9^\circ$), both streams operate on the shared fused representation $f_{av}$. Their gradients show moderate correlation. After the first cross-stream interaction, the angle increases to $81.2^\circ$. This indicates that cross-attention does not mix representations. Instead, it encourages divergence between the two streams. After the second SAL layer, the angle further increases to $86.1^\circ$. This is close to orthogonality. It shows that the two streams have formed distinct functional roles. Their optimization mainly occurs in different subspaces. In the final stage, the angle slightly decreases to $82.7^\circ$. This reflects limited alignment before final fusion, while maintaining relative independence.

Importantly, the gradient angles do not follow a random distribution. They increase monotonically from moderate correlation to a high stable range. This behavior is induced by the model structure and reflects structured representation separation, rather than a random effect in high-dimensional space.

\subsection{Serial vs. Parallel Dual-Stream Comparison}
To directly show the performance gain from reducing gradient interference, we compare the dual-stream design under serial and parallel settings.
In the serial setting, the two streams are connected sequentially. The output of the first stream is used as the input of the second stream. In the parallel setting, we use the proposed decoupled framework. The two structures are identical except for the connection pattern.

The results are shown in Table~\ref{tab:ablation_parallel}. The serial structure achieves $95.1$ mAP. The parallel decoupled structure improves the performance by $0.5$ and reaches $95.6$ mAP. This gain is achieved on a strong baseline. This result supports the analysis in Section~\ref{sec:assum}. 
In the serial structure, the second stream operates on features that have already been transformed by the first stream. This leads to repeated representation over a biased feature distribution. Such deep coupling introduces strong gradient interference. It prevents the model from reaching optimal solutions for both ITC and ISR.
In contrast, the parallel structure separates the two streams. The ITC-Stream focuses on low-frequency and smooth temporal modeling. The ISR-Stream focuses on high-frequency and discriminative relational reasoning. During backpropagation, their gradients do not interfere with each other.
This structural separation reduces the seesaw effect. Each stream can converge within its own functional space. This leads to better overall performance.

\section{Comparative Experimental Analysis}

To comprehensively evaluate the effectiveness and advancement of the proposed D$^2$Stream, we conduct experiments on two authoritative benchmarks: AVA-ActiveSpeaker~\cite{AVA} and Columbia ASD~\cite{Columbia}.

\noindent\textbf{AVA-ActiveSpeaker:} Derived from Hollywood movies, it provides 1–10s face-track utterances evaluated by mAP, with challenges including diverse languages, poor audio-visual quality and asynchronization.

\noindent\textbf{Columbia ASD:} As a standard ASD benchmark, it consists of an 87-minute panel discussion with 5 alternating speakers, where 2–3 are visible at any time.

\subsection{Results on AVA-ActiveSpeaker}

The quantitative results on AVA-ActiveSpeaker are shown in Table~\ref{tab:big_ava}. Since TalkNCE (ICASSP 2024) achieved 95.5 mAP, later methods such as LASER (WACV 2026) have not surpassed this result. The field has reached a clear performance plateau. D$^2$Stream achieves 95.6\% mAP and surpasses TalkNCE. It sets a new state-of-the-art and breaks the long-standing performance limit.

From the efficiency perspective, D$^2$Stream uses only 0.54 GFLOPs and 24.3M parameters. Compared to AFs-Net, it improves accuracy by 0.2 while reducing computation by about 79\%. Compared to TalkNCE, it uses about 10M fewer parameters with similar FLOPs. It achieves better accuracy and efficiency at the same time. These results show that decoupling ITC (Intra-speaker Temporal Continuity) and ISR (Inter-personal Social Relation) reduces optimization conflict. It allows better use of complementary information and improves performance under lightweight constraints.

\begin{table*}[!t]
  \centering
  \caption{Performance comparison on the AVA-ActiveSpeaker dataset~\cite{AVA}. Our \textsc{D$^{2}$Stream} achieves state-of-the-art performance with the highest mAP of $95.60\%$, while maintaining competitive computational efficiency.}
  \begin{tabular}{@{}lcccccc@{}}
    \TopRule
    Method & Candidate Type & Venue & FLOPs(G) & Params(M) & mAP(\%) \\
    \midrule
    TalkNet~\cite{TalkNet}        & Single   & ACM MM 2021 & $0.51$ & $15.7$ & $92.3$ \\
    Sync-TalkNet~\cite{Sync-TalkNet} & Single & MLSP 2022   & $1.6$  & $14.6$ & $89.8$ \\
    ASD-Trans~\cite{ASD-Trans}    & Single   & ICASSP 2022 & $0.55$ & $14.9$ & $93.0$ \\
    ADENet~\cite{ADENet}          & Single   & TMM 2022    & $22.8$ & $33.2$ & $93.2$ \\
    Light-ASD~\cite{Light-ASD}    & Single   & CVPR 2023   & $0.20$ & $1.02$ & $94.1$ \\
    LR-ASD~\cite{LR-ASD}          & Single   & IJCV 2025   & $0.51$ & $0.84$ & $94.5$ \\
    SCAN~\cite{SCAN}              & Single   & ICASSP 2025 & -- & -- & $94.0$ \\
    GateFusion~\cite{GateFusion}  & Single   & WACV 2026   & -- & -- & $95.0$ \\
    \midrule
    MAAS~\cite{MAAS}              & Multiple & ICCV 2021   & $1.6$  & $23.0$ & $88.8$ \\
    UniCon~\cite{Unicon}          & Multiple & ACM MM 2021 & $3.0$  & $23.8$ & $92.2$ \\
    ASDNet~\cite{ASDNet}          & Multiple & ICCV 2021   & $13.2$ & $51.0$ & $93.5$ \\
    EASEE~\cite{EASEE}            & Multiple & ECCV 2022   & $4.3$  & $26.8$ & $94.1$ \\
    SPELL+~\cite{SPELL+}          & Multiple & ECCV 2022   & $19.6$ & $51.2$ & $94.9$ \\
    LoCoNet~\cite{LocoNet}        & Multiple & CVPR 2024   & $0.51$ & $34.3$ & $95.2$ \\
    \rowcolor{blue!5}
    TalkNCE~\cite{TalkNCE}       & Multiple & ICASSP 2024 & $0.51$ & $34.3$ & $\underline{95.5}$ \\
    AFs-Net~\cite{AFs-Net}        & Multiple & ICASSP 2025 & $2.6$  & $18.9$ & $95.4$ \\
    LASER~\cite{LASER}            & Multiple   & WACV 2026   & -- & -- & $95.4$ \\
    \rowcolor{yellow!5}
    \textbf{D$^2$Stream (ours)}   & Multiple & --          & $0.54^{\textcolor{red}{\scriptsize \uparrow 0.03}}$ & $24.3^{\textcolor{blue}{\scriptsize \downarrow 10.0}}$ & $\mathbf{95.6}^{\textcolor{red}{\scriptsize \uparrow 0.1}}$ \\
    \bottomrule
  \end{tabular}
  \label{tab:big_ava}
\end{table*}

\subsection{Results on Columbia ASD}

To further evaluate cross-speaker generalization, we compare with existing methods on Columbia ASD. The results are shown in Table~\ref{tab:columbia}. D$^2$Stream achieves an average accuracy of 81.5\%. It outperforms AFs-Net (80.4\%) and achieves the best overall result. For individual categories, D$^2$Stream shows strong performance on challenging speakers such as Lieb (90.0\%) and Long (87.7\%). It also maintains competitive results on the other categories. These results further demonstrate the robustness and state-of-the-art performance of the proposed method.


\begin{table*}[!t]
  \centering
  \caption{Performance comparison on the Columbia ASD dataset~\cite{Columbia} across five test speakers. \textsc{D$^{2}$Stream} achieves the highest average score.}
  \begin{tabular}{@{}ll ccccc c@{}}
    \toprule
    \multirow{2}{*}{Venue} & \multirow{2}{*}{Method} & \multicolumn{5}{c}{Speaker} & \multirow{2}{*}{Avg} \\ \cmidrule(lr){3-7}
    & & Lieb & Long & Bell & Boll & Sick & \\
    \midrule
    ACM MM 2021 & TalkNet (2021)       & $68.7$ & $43.8$ & $43.6$ & $66.6$ & $58.1$ & $56.2$ \\
    CVPR 2023   & Light-ASD (2023)     & $87$   & $74.5$ & $82.7$ & $75.7$ & $85.4$ & $\underline{81.1}$ \\
    CVPR 2024   & LoCoNet (2023)       & $80.2$ & $80.4$ & $54$   & $49.1$ & $76.8$ & $68.1$ \\
    ICASSP 2025 & AFs-Net (2025)       & $85.9$ & $81.1$ & $73.5$ & $77$   & $84.7$ & $80.4$ \\
    \rowcolor{yellow!5}
    --          & \textbf{D$^2$Stream (ours)} & $90$   & $87.7$ & $71.5$ & $76.7$ & $81.4$ & $\mathbf{81.5}$ \\
    \bottomrule
  \end{tabular}
  \label{tab:columbia}
\end{table*}

\section{Ablation Study}
We also quantitatively evaluate the independent contributions of each core module in the model design through a series of ablation experiments. These modules include modal input, context-aware gating, dual-stream branches, and the number of decoupled interaction layers.

\begin{table*}[!t]
  \centering
  \caption{Overall ablation studies on our method.
    (a) Modality contribution.
    (b) Context-aware gating mechanism.
    (c) Temporal and speaker interaction branches.
    (d) Number of decoupled dual-stream interaction layers.
  }
  \begin{tabular}{@{}l cccccc@{}}
    \toprule
    \textbf{Ablation Setting} & FLOPs(G) & $\Delta$FLOPs(G) & Params(M) & $\Delta$Params(M) & mAP(\%) & $\Delta$mAP(\%) \\
    \midrule
    \rowcolor{yellow!5}
    Full Model                & $0.54$   & --               & $24.34$   & --                & $95.6$  & --             \\
    \rowcolor{gray!5}
    \multicolumn{7}{c}{\textcolor{black}{\textbf{(a) Modality Ablation}}} \\
    w/o Visual Features       & $0.04$   & $-0.50$          & $10.99$   & $-13.35$          & $51.3$  & $-44.3$        \\ 
    w/o Audio Features        & $0.51$   & $-0.03$          & $18.99$   & $-5.35$           & $85.2$  & $-10.4$        \\
    \rowcolor{gray!5}
    \multicolumn{7}{c}{\textcolor{black}{\textbf{(b) Context-Aware Gating Ablation}}} \\
    w/o Gate                  & $0.54$   & $0$              & $24.18$   & $-0.16$           & $95.4$  & $-0.2$         \\
    \rowcolor{gray!5}
    \multicolumn{7}{c}{\textcolor{black}{\textbf{(c) Dual-Stream Ablation}}} \\
    base                      & $0.53$   & $-0.01$          & $18.81$   & $-5.53$           & $89.1$  & $-6.5$         \\
    w/o ITC-Stream            & $0.54$   & $0$              & $20.52$   & $-3.82$           & $94.0$  & $-1.6$         \\
    w/o ISR-Stream            & $0.54$   & $0$              & $20.52$   & $-3.82$           & $94.5$  & $-1.1$         \\
    \rowcolor{gray!5}
    \multicolumn{7}{c}{\textcolor{black}{\textbf{(d) Number of Decoupled Dual-Stream Interaction Layers}}} \\
    $1$ Layer                 & $0.54$   & $0$              & $21.58$   & $-2.76$           & $95.2$  & $-0.4$         \\
    $2$ Layers (Selected)     & $0.54$   & --               & $24.34$   & --                & $95.6$  & --             \\
    $3$ Layers                & $0.55$   & $+0.01$          & $27.10$   & $+2.76$           & $95.3$  & $-0.3$         \\
    \bottomrule
  \end{tabular}
  \label{tab:ablation_all}
\end{table*}

\subsection{Modality Ablation}

To study the respective contributions of visual and audio modal information to the final prediction, we replace the feature vector of the ablated modal with a zero vector of the same dimension. This ensures that the model parameter count and data flow remain basically consistent while observing the impact of a single variable on prediction accuracy.


The ablation results in Table~\ref{tab:ablation_all}(a) show that both visual and audio modalities have irreplaceable independent contributions in the overall framework. 

Removing the visual branch reduces FLOPs by 0.48 G and the parameter count by 13.36 M. At the same time, mAP drops sharply from 95.5 to 51.3. This highlights the core supporting role of visual features in speech-lip synchronization in multi-speaker scenarios.

Removing the audio branch reduces FLOPs by 0.01 G and the parameter count by 5.35 M, with mAP dropping to 85.2. This shows the key value of audio information in characterizing the temporal continuity of speech. 

Overall, the two modalities provide complementary information from spatial-temporal cues and temporal dynamics respectively. The absence of either will cause significant performance degradation.

\subsection{Context-Aware Gating Ablation}

We introduce a context-aware gating module in the fusion stage to adaptively filter cross-modal features. To evaluate its effect, we remove the gating mechanism and directly concatenate $\tilde{f}_a$ and $\tilde{f}_v$ as the fused feature $f_{av}$.

The experimental results in Table~\ref{tab:ablation_all}(b) show that introducing context-aware gating further improves the model's mAP from 95.4 to 95.6 on the AVA-ActiveSpeaker validation set, while only introducing negligible computational load. This performance gain strongly proves the necessity of the secondary modulation mechanism in complex audio-visual scenarios.

\subsection{Dual-Stream Ablation}

To investigate the contributions of the ITC-Stream and ISR-Stream, we perform an ablation study by removing the corresponding stream, following the same setting as in Section~\ref{sec:assum}.

\begin{figure}[t]
  \centering
  \includegraphics[width=0.5\textwidth]{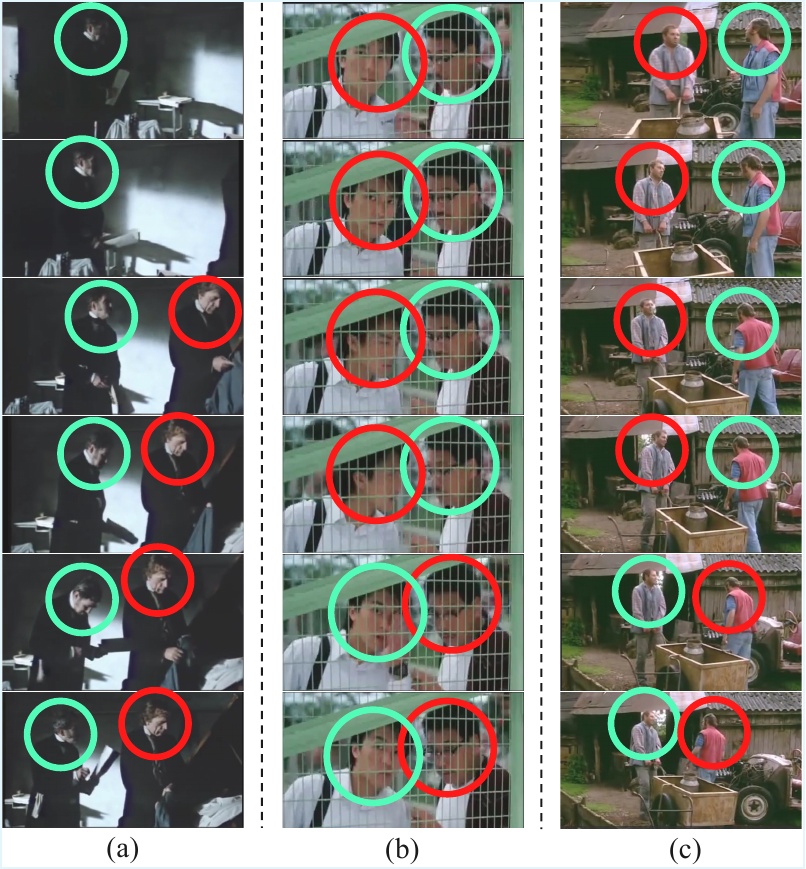}
  \caption{Case study in challenging scenarios. Our full model effectively handles complex cases via decoupled modeling: (a) Lighting and multi-person interference: precisely aligning multimodal features under facial shadows to avoid false positives caused by visual clarity bias; (b) Partial mouth occlusion: leveraging ITC to capture temporal continuity, ensuring stable predictions during occlusions; (c) Extreme poses (back to camera): utilizing ISR to perceive social context and interaction focus.}
  \label{fig:dxt}
\end{figure}

As shown in Table~\ref{tab:ablation_all}(c), both streams contribute significantly to performance and play complementary roles that are indispensable. Removing the ITC-Stream decreases mAP from 95.6 to 94.0 (-1.6), while removing the ISR-Stream reduces mAP from 95.6 to 94.5 (-1.1). Notably, the base model already achieves a strong baseline of 89.1 mAP, mainly because the AVA-ActiveSpeaker dataset contains many single-speaker clips that can be handled by simple multimodal fusion. The core value of the dual-stream structure lies in solving complex multi-person scenarios beyond single-speaker cases. We further conduct a case study on samples where the full model outperforms all ablated variants, as illustrated in Fig.~\ref{fig:dxt}, verifying the importance of both branches for accurate speaker detection in challenging scenes. More analyses are provided in the appendix.

\subsection{Number of Decoupled Dual-Stream Interaction Layers}
We conduct an ablation study to analyze how the number of dual-stream interaction layers affects model performance.

As reported in Table~\ref{tab:ablation_all}(d), performance first increases and then decreases with the number of layers. The two-layer structure achieves the best result (95.6 mAP), while the one-layer and three-layer counterparts obtain 95.2 and 95.3 mAP respectively. A single interaction layer only supports one-step feature alignment and fails to fully exploit complementary information between heterogeneous representations. The two-layer design allows each type of feature to evolve sufficiently in its own subspace before information fusion, enabling stable collaborative modeling, which aligns with the observed gradient divergence and stabilization trend in Section~\ref{sec:gaa}. Increasing to three layers expands model capacity but introduces redundant information mixing and overfitting, leading to performance degradation.

\section{Conclusion}
From the perspective of representation learning, we re-examine the essential structure of the audio-visual speaker detection task. We point out the inherent conflict in inductive bias between ITC modeling and ISR modeling.

Based on this observation, we propose D$^2$Stream, a simple and elegant decoupled dual-stream framework. It explicitly separates the two functions in structure, allowing them to be optimized in independent subspaces. Effective collaboration is achieved through controlled cross-stream interaction.

Systematic experimental analysis shows that this design alleviates gradient interference in the shared parameter space. It breaks the long-standing performance bottleneck of this task while keeping the model lightweight. It achieves leading experimental results on both the AVA-ActiveSpeaker and Columbia ASD datasets.

In the future, we plan to adapt D$^2$Stream to real-time streaming processing systems. This is to verify its application potential in real-time industrial scenarios such as intelligent conferences and human-computer interaction.


\bibliographystyle{ACM-Reference-Format}
\bibliography{strings,appendix/app}

\appendix

\clearpage
\onecolumn
\title{Appendix of Dual-Stream Decoupled Learning for Temporal Consistency and Speaker Interaction in AVSD}

\section{Where Baselines Fail, Ours Prevails}
\label{app:sec:exp-A}
Through sample analysis, we observe that our method achieves consistent advantages over existing models in the following four typical challenging scenarios for Audio-Visual Speaker Detection (AVSD). The superior performance in these complex cases constitutes the primary driver of the overall performance gains by improving robustness in failure-prone scenarios.

\subsection{Blurred or Missing Facial Visual Information}
\label{app:sec:exp-A-1}
\begin{figure}[h]
\centering
\includegraphics[width=0.85\textwidth]{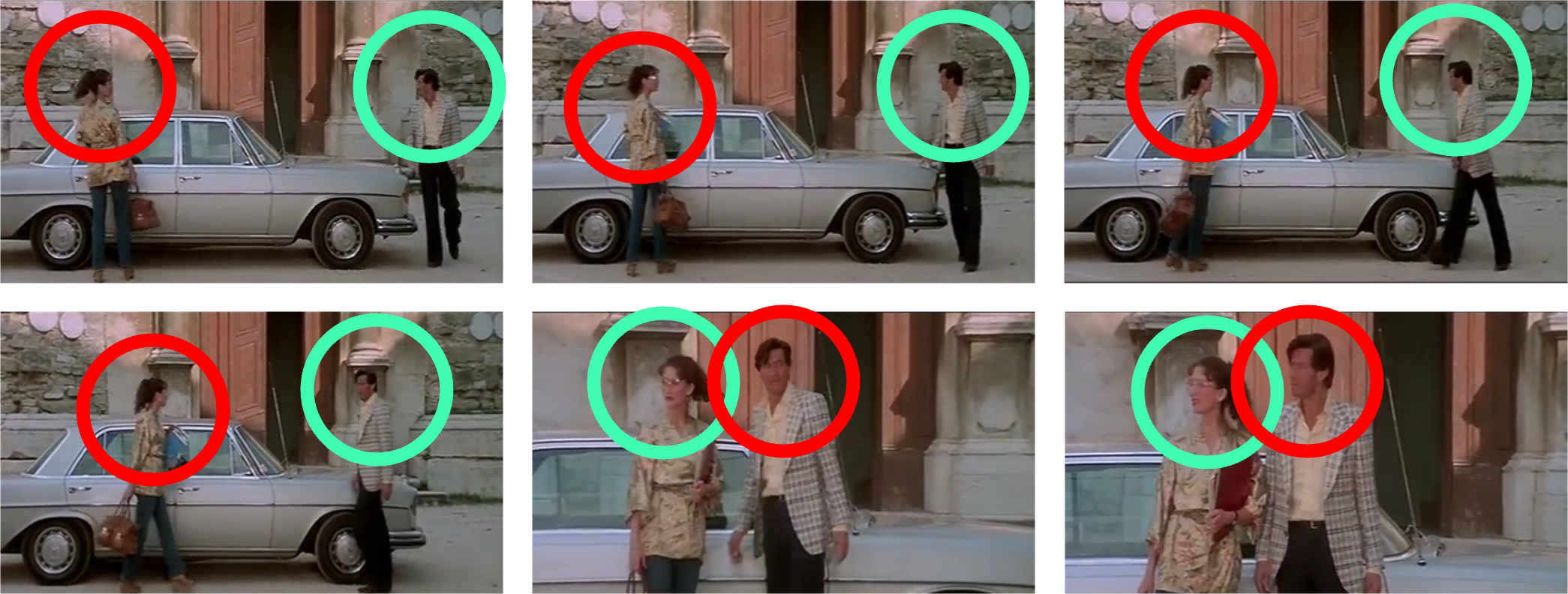}
\caption{Scenario of blurred or missing facial visual information.}
\label{fig:blurface}
\end{figure}

As shown in Fig.~\ref{fig:blurface}, this scenario is characterized by low resolution and non-frontal viewpoints. Under such long-shot shooting conditions, the key head regions of target speakers occupy only a very limited number of pixels in the entire image, leading to severe degradation or loss of fine-grained facial features. Especially in the first three frames of the sequence, both candidate speakers appear in profile view, making lip motion cues largely unobservable. Traditional models heavily rely on these features. In such extreme cases lacking visual cues, existing methods often fail due to the inability to extract effective discriminative visual embeddings.

\begin{figure}[h]
\centering
\includegraphics[width=0.85\textwidth]{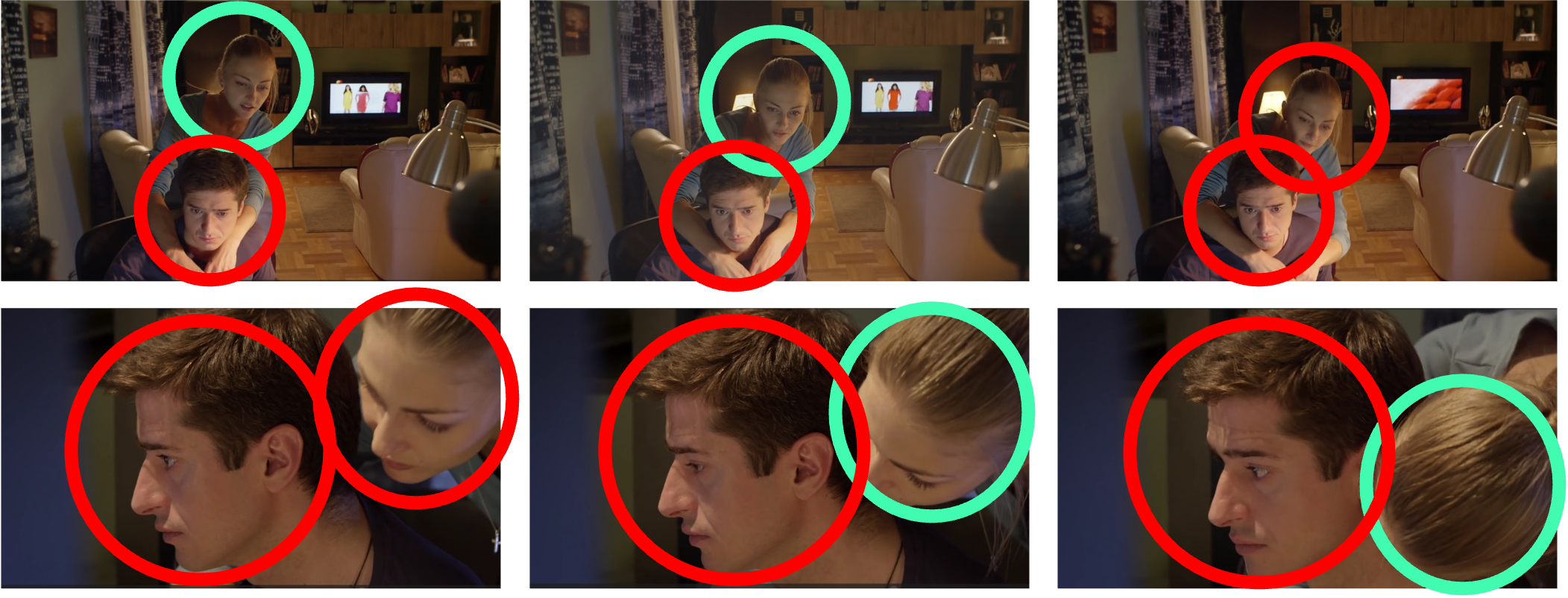}
\caption{Scenario of sudden viewpoint transition.}
\label{fig:viewchange}
\end{figure}

\subsection{Sudden Viewpoint Transition}
\label{app:sec:exp-A-2}
As shown in Fig.~\ref{fig:viewchange}, this sequence exhibits an sudden viewpoint transition from panoramic view to close-up. The core challenge of such scenes lies in the severe temporal discontinuity of visual semantic features: drastic scaling of the frame causes significant shifts of speaker appearance representations in the feature space, which further breaks the model’s long-range dependency modeling of visual context. Existing methods typically rely heavily on inter-frame feature continuityD2; under such abrupt changes, they easily suffer from feature alignment failure, reflected in obvious fluctuations in predictions.

\subsection{Multi-source Environmental Interference}
\label{app:sec:exp-A-3}
\begin{figure}[h]
\centering
\includegraphics[width=0.85\textwidth]{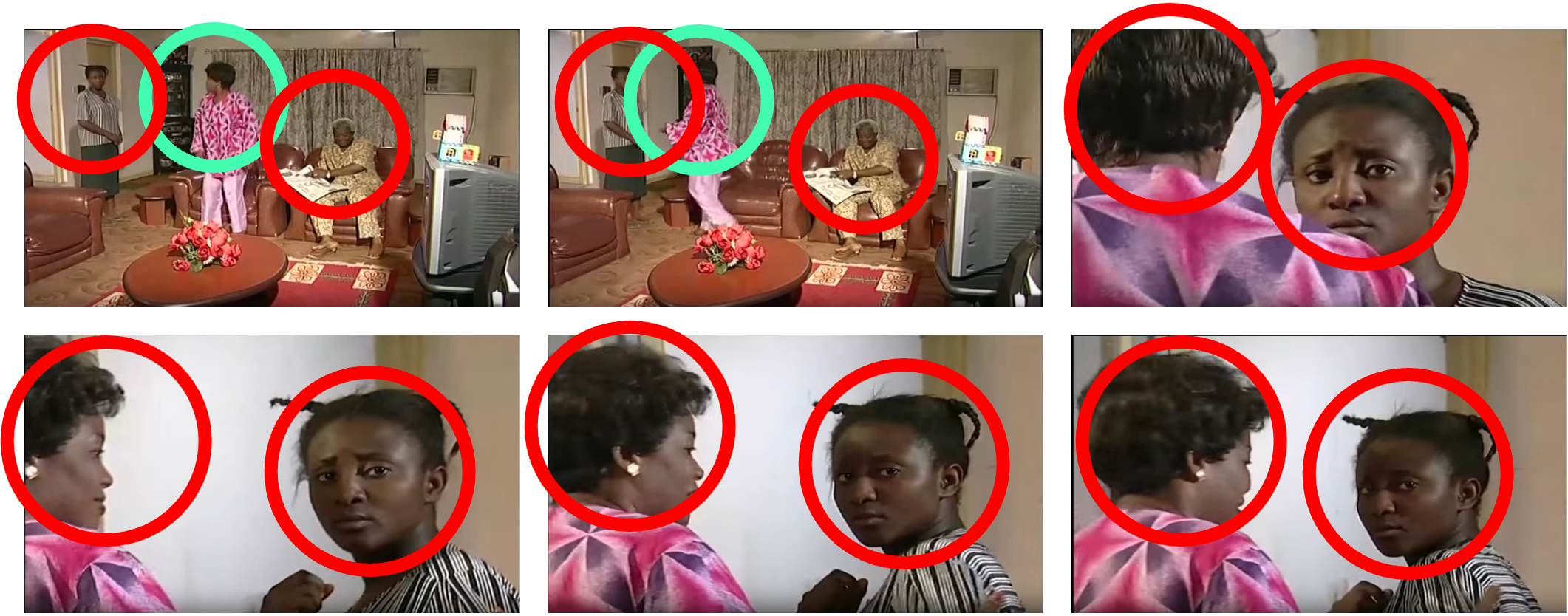}
\caption{Scenario of multi-source environmental interference.}
\label{fig:multisource}
\end{figure}

As shown in Fig.~\ref{fig:multisource}, this sample contains two types of challenges: visual ambiguity and background audio interference. In the first two frames, the real speaker utters only briefly, with a heavily blurred face shown mostly in profile, requiring the model to accurately identify the speaker among three candidates. In the following four frames, strong background music is added, corrupting audio cues. Meanwhile, a woman in purple appears at the center of the frame with clear facial features and slight lip movements, yet she is not actually speaking. In this case, the model must not only suppress misleading background noise but also jointly reason based on facial states, lip motion consistency, and voice activity features.

\begin{figure}[h]
\centering
\includegraphics[width=0.85\textwidth]{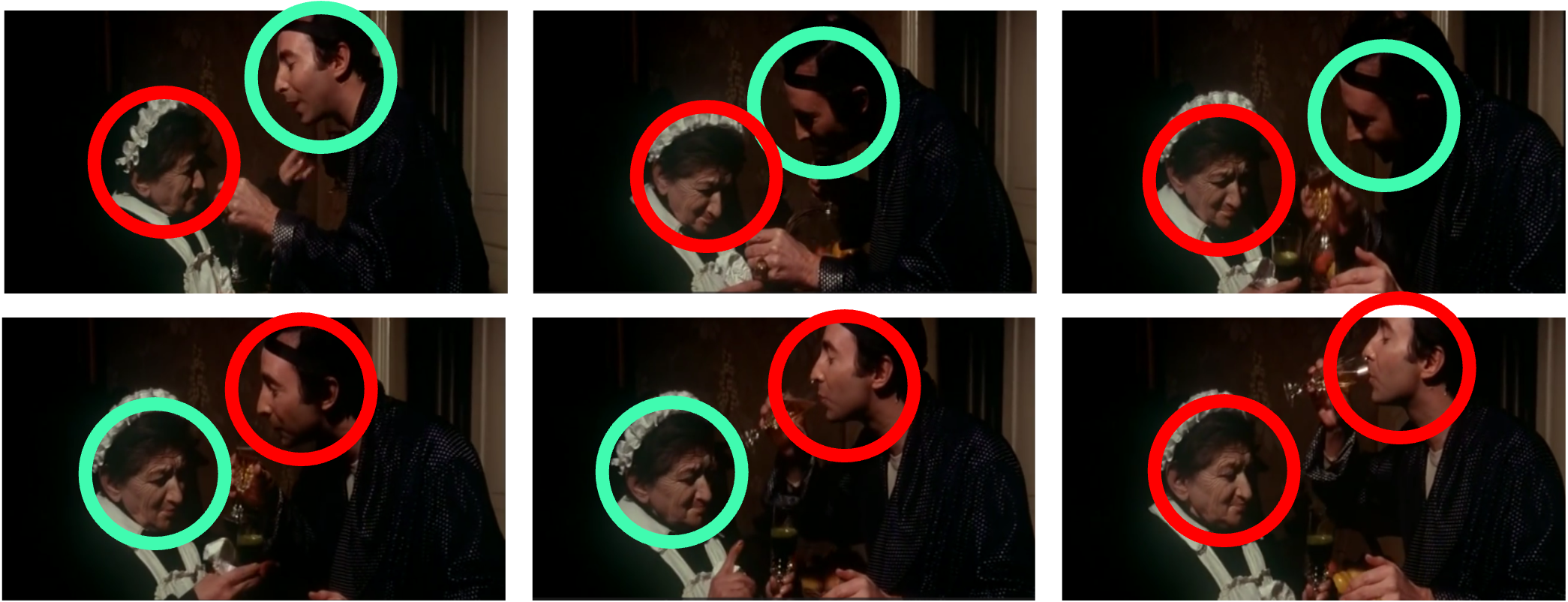}
\caption{Scenario of competitive interference between short responses and long monologues.}
\label{fig:shortresponse}
\end{figure}

\subsection{Competitive Interference Between Short Responses and Long Monologues}
\label{app:sec:exp-A-4}
As shown in Fig.~\ref{fig:shortresponse}, the person on the right speaks continuously for a long time, while the person on the left only produces a short response of about 0.3 seconds in the middle. Since this response is extremely brief, accompanied by inconspicuous lip motion and weak audio energy, the model can easily be dominated by the continuous speaker on the right and thus ignore this low-saliency, short-duration speech event.

\section{Qualitative Ablation Analysis}
\label{app:sec:exp-B}
To further analyze the source of performance gains, we conduct ablation studies on the four challenging samples detailed in Section~\ref{app:sec:exp-A}, comparing the prediction outputs of the full model and its single-stream variants. Here, \texttt{Only\_ITC} denotes the variant retaining only the ITC-Stream, and \texttt{Only\_ISR} denotes the variant retaining only the ISR-Stream. We focus on the respective contribution boundaries of the two streams and verify that the dual-stream interaction indeed provides complementary performance gains.

\subsection{Blurred or Missing Facial Visual Information}
As can be seen from Table~\ref{tab:male}, in scenarios with blurred or missing facial visual information, the full model yields higher responses on the four positive samples than both single-stream variants, while maintaining zero responses on the two negative samples. Although \texttt{Only\_ITC} and \texttt{Only\_ISR} can produce weak responses on some frames, their scores are significantly lower when used alone. This indicates that neither temporal continuity nor intra-frame relations alone are sufficient for stable discrimination under degraded visual conditions. In contrast, the full model preserves both inter-frame continuous evidence and intra-frame relational cues simultaneously.

\begin{table}[h]
\centering
\caption{Ablation analysis on the scenario of blurred or missing facial visual information, focusing on the male key speaker candidate (from Section~\ref{app:sec:exp-A-1}).}
\label{tab:male}
\begin{tabular}{lcccccc}
\toprule
             & Frame 1 & Frame 2 & Frame 3 & Frame 4 & Frame 5 & Frame 6 \\
\midrule
GroundTruth  & \textcolor{green}{1}    & \textcolor{green}{1}    & \textcolor{green}{1}    & \textcolor{green}{1}    & \textcolor{red}{0}    & \textcolor{red}{0}    \\
Ours         & 0.42 & 0.37 & 0.56 & 0.53 & 0.00    & 0.00    \\
Only\_ITC    & 0.13 & 0.10 & 0.09 & 0.10 & 0.00    & 0.00    \\
Only\_ISR    & 0.05 & 0.07 & 0.04 & 0.04 & 0.00    & 0.00    \\
\bottomrule
\end{tabular}
\end{table}

\subsection{Sudden Viewpoint Transition}
As shown in Table~\ref{tab:female}, in the sudden viewpoint transition scenario, \texttt{Only\_ITC} achieves relatively high responses in the first two frames (0.96 / 0.91), but its confidence drops notably in the later positive segments (0.32 / 0.35). This indicates that while temporal continuity helps preserve historical evidence, it struggles to adapt to abrupt appearance changes. 
\texttt{Only\_ISR} performs reasonably in the initial frames (0.92 / 0.90), yet produces weaker responses in the later positive frames (0.16 / 0.20), suggesting that relying solely on intra-frame relations is insufficient to re-establish stable predictions after viewpoint shifts. In contrast, the full model maintains the highest confidence in the first two frames (0.99 / 0.96) and significantly outperforms both single-stream variants in the later positive frames (0.68 / 0.56). This demonstrates that the ITC-Stream preserves temporal continuity, while the ISR-Stream complements discriminative cues after appearance changes. Their synergy effectively mitigates feature shifts caused by viewpoint transitions, resulting in more robust predictions.

\begin{table}[h]
\centering
\caption{Ablation analysis on the scenario of Sudden Viewpoint Transition, focusing on the female key speaker candidate (from Section~\ref{app:sec:exp-A-2}).}
\label{tab:female}
\begin{tabular}{lcccccc}
\toprule
             & Frame 1 & Frame 2 & Frame 3 & Frame 4 & Frame 5 & Frame 6 \\
\midrule
GroundTruth  & \textcolor{green}{1}    & \textcolor{green}{1}    & \textcolor{red}{0}    & \textcolor{red}{0}    & \textcolor{green}{1}    & \textcolor{green}{1}    \\
Ours        & 0.99 & 0.96 & 0.00 & 0.00 & 0.68 & 0.56 \\
Only\_ITC   & 0.96 & 0.91 & 0.00 & 0.00 & 0.32 & 0.35 \\
Only\_ISR   & 0.92 & 0.90 & 0.01 & 0.00 & 0.16 & 0.20 \\
\bottomrule
\end{tabular}
\end{table}

\subsection{Multi-source Environmental Interference}
Table~\ref{tab:purple} clearly reflects the different roles of the two streams under multi-source environmental interference. \texttt{Only\_ITC} still yields relatively high responses on the first two positive frames while maintaining low scores on negative frames, indicating its strength in maintaining stable judgments using temporal trajectories. \texttt{Only\_ISR} shows obvious false activations on later negative frames, demonstrating that relying solely on intra-frame relations makes the model susceptible to visual centrality or local motions, misclassifying non-speakers as active speakers. The full model maintains strong responses on positive frames while suppressing negative frames to near zero, showing that dual-stream interaction preserves both temporal evidence and relational constraints simultaneously, thus reducing the bias of single-stream variants.

\begin{table}[h]
\centering
\caption{Ablation analysis on the scenario of multi-source environmental interference, focusing on the purple-clothed key speaker candidate (from Section~\ref{app:sec:exp-A-3}).}
\label{tab:purple}
\begin{tabular}{lcccccc}
\toprule
             & Frame 1 & Frame 2 & Frame 3 & Frame 4 & Frame 5 & Frame 6 \\
\midrule
GroundTruth  & \textcolor{green}{1}    & \textcolor{green}{1}    & \textcolor{red}{0}    & \textcolor{red}{0}    & \textcolor{red}{0}    & \textcolor{red}{0}    \\
Ours         & 0.52 & 0.30 & 0.00 & 0.00 & 0.01 & 0.00 \\
Only\_ITC    & 0.25 & 0.11 & 0.01 & 0.03 & 0.04 & 0.03 \\
Only\_ISR    & 0.04 & 0.04 & 0.05 & 0.44 & 0.35 & 0.35 \\
\bottomrule
\end{tabular}
\end{table}

\subsection{Competitive Interference Between Short Responses and Long Monologues}
Table~\ref{tab:left} provides a direct comparison on the short-response scenario. \texttt{Only\_ISR} can hardly detect this extremely brief utterance, whereas \texttt{Only\_ITC} already produces noticeable responses on response frames, indicating that short speech events mainly rely on temporal continuity for capture. The full model further elevates scores on frames 4 and 5, showing that the ISR-Stream does not perform short detection independently, but interacts with the ITC-Stream to help the model distinguish between the two adjacent but distinct speaking behaviors: ``sustained monologue'' and ``short response''.

\begin{table}[h]
\centering
\caption{Ablation analysis on the scenario of competitive interference between short responses and long monologues, focusing on the left-side key speaker candidate (from Section~\ref{app:sec:exp-A-4}).}
\label{tab:left}
\begin{tabular}{lcccccc}
\toprule
             & Frame 1 & Frame 2 & Frame 3 & Frame 4 & Frame 5 & Frame 6 \\
\midrule
GroundTruth  & \textcolor{red}{0}    & \textcolor{red}{0}    & \textcolor{red}{0}    & \textcolor{green}{1}    & \textcolor{green}{1}    & \textcolor{red}{0}    \\
Ours         & 0.00    & 0.00    & 0.00 & 0.52 & 0.81 & 0.06 \\
Only\_ITC    & 0.00    & 0.00    & 0.02 & 0.28 & 0.52 & 0.02 \\
Only\_ISR    & 0.00    & 0.00    & 0.01 & 0.04 & 0.03 & 0.01 \\
\bottomrule
\end{tabular}
\end{table}

\section{Why We Do Not Use Dual-Stream Losses}
\label{app:sec:exp-C}
Inspired by LoCoNet~\cite{LocoNet}, which improves performance by applying independent supervision to two modality encoders, we further attempt to introduce additional branch-level losses for the dual-stream branches. Specifically, we add separate frame-level supervision to the ITC-Stream and ISR-Stream respectively, and optimize them jointly with the final fused prediction. As shown in Table~\ref{tab:dual_loss}, experimental results show that this setup does not yield additional performance improvements; instead, model performance slightly drops from 95.6 to 95.5.

\begin{table}[h]
\centering
\caption{Ablation on dual-stream independent losses.}
\label{tab:dual_loss}
\begin{tabular}{lcc}
\toprule
Setting                        & AVA mAP & $\Delta$ \\
\midrule
D$^2$Stream (full)             & 95.6    & - \\
+ Dual-stream independent losses & 95.5    & \textcolor{red}{-0.1} \\
\bottomrule
\end{tabular}
\end{table}

This result indicates that the dual-stream branches in our work are fundamentally different from the audio and visual encoders in LoCoNet. The two branches in LoCoNet correspond to two physically independent modalities, and separate supervision helps enhance their respective discriminability. In contrast, both the ITC-Stream and ISR-Stream in this paper are built upon the same fused representation. Their role is not to repeatedly perform the same classification task, but to extract inter-frame temporal consistency and intra-frame relation information separately, which complement each other in subsequent cross-stream interaction. Applying strong supervision to both streams individually imposes the same label constraint on two intermediate representations simultaneously, which tends to force the two streams to converge toward similar discriminative directions prematurely, weakening their specialization in independent subspaces and the room for interaction.

Therefore, we adopt main supervision on the fusion branch, combined with unimodal auxiliary supervision and contrastive constraints to stabilize representation learning, without imposing additional classification losses on the intermediate dual-stream branches.

\section{Implementation Details} \label{app:sec:details}

\subsection{Visual Encoder}
The visual encoder adopts a three-level feature encoding architecture, consisting of a 3D convolutional layer, ResNet-18 \cite{resnet}, and a Visual Temporal Convolutional Network (V-TCN) \cite{vtcn}. A 3D convolutional layer first extracts spatial-temporal features from the input face frame sequences, and the extracted features are then fed into the ResNet-18 backbone composed of 4 residual blocks to increase the channel dimension. Finally, V-TCN, which is stacked with 5 depth-wise separable convolution blocks, models temporal dependencies to complete temporal modeling and dimension adaptation of visual features.

\subsection{Audio Encoder}
The audio encoder is built based on the VGGFrame structure proposed in LoCoNet, with weights initialized using the VGGish~\cite{vggish} model pretrained on the AudioSet dataset~\cite{audioset}. It retains the core convolutional backbone of VGGish while removing its original preprocessing and post-processing modules. Input audio is resampled to 16 kHz uniformly, and 13-dimensional Mel Frequency Cepstral Coefficients (MFCC) are extracted as basic features, which are then encoded through multiple convolution-batch normalization-ReLU units and dimensionally reduced by adaptive average pooling. The final output is 128-dimensional temporal audio features, which are consistent with the dimension of visual features.

\subsection{Data Preprocessing and Augmentation}

\noindent\textbf{Visual Data:} Cropped face regions are uniformly resized to 112×112 grayscale images. Visual data augmentation strategies include random size cropping, horizontal flipping, and random rotation, and zero matrices are used for padding missing face frames to ensure consistent sequence length.

\noindent\textbf{Audio Data:} Audio data augmentation adopts a noise superposition strategy: audio segments from non-current samples in the training set are randomly selected as noise sources, and the noise audio is superimposed on the target audio with a Signal-to-Noise Ratio (SNR) of -5 dB to 5 dB. Loop padding is used to complement the noise audio if its length is insufficient, ensuring the audio feature length is aligned with the time steps of the visual frame sequence.

\subsection{Model Component}
The multi-head self-attention mechanism adopted in this paper is configured with 8 attention heads.

\subsection{Training Configuration}
Experiments are implemented based on the PyTorch framework, and the model is deployed on 4 RTX 2080Ti GPUs with a distributed training strategy to ensure the consistency of multi-GPU training. The Adam optimizer \cite{adam} is selected, with an initial learning rate of $5 \times 10^{-5}$ that decays by 5\% per epoch, and the model is trained with a batch size of 4 for a total of 35 epochs.

\end{document}